# The Improvement of the Air Quality due to Traffic Halting in Los Angeles and Potential Health Care Risk during the COVID-19 Outbreak

Jiani Yang [a,*], Yuan Wang [a,b], Joseph Pinto [c], Le Kuai [a,b,d], King-Fai Li [e], Stanley P Sander [a,b], Yuk L Yung [a,b]

[a] Division of Geological and Planetary Sciences, California Institute of Technology, Pasadena, CA, 91125 USA

[b] Jet Propulsion Laboratory, 4800 Oak Grove Dr, Pasadena, CA 91109 USA

c Environmental Sciences and Engineering, Gillings School of Global Public Health, University of North Carolina at Chapel Hill, 135 Dauer Drive, Chapel Hill, NC 27599 USA

d Joint Institute for Regional Earth System Science and Engineering, University of California, Los Angeles, CA 90095 USA

e Department of Environmental Sciences, University of California, Riverside, 1140 Batchelor Hall Riverside, CA 92521 USA

*  corresponding to yjn@caltech.edu

**Abstract**

**Background**: On March 19 2020, the government of California ordered all 40 million Californians to stay at home in the coming weeks as the result of the escalation of the coronavirus disease 2019 (COVID-19) pandemic. As lockdowns were implemented, the significant changes caused by these restrictions brought the dramatic improvement in air quality in metropolitan cities such as Los Angeles (LA Basin).

**Methods**: We use real time data from The South Coast Air Quality Management District (South Coast AQMD), and California Department of Transportation (PeMS) to evaluate the anthropogenic drivers of the pollution sources. We fit the regression analysis to compare the correlation of 7 variables including traffic flow, truck flow, traffic speed, $NO_2$, CO, $PM_{2.5}$ and $O_3$. We also mapped the monthly spatial variation and hourly heatmap of those 7 variables in 2020 to understand the impacts of the lockdown on different locations in LA Basin.

**Results**: In the Los Angeles Basin, the traffic flow on highways started to drop a intensely by 20.86 % when initiating the stay at home order and it continued decreasing by 32.92% the first week, 30.94% the second week, 37.9% the third week, and 33.57% at the fourth week following the lockdown compared to corresponding dates during the year of 2019. The average truck flow for each sensor is generally higher in 2020 than 2019 before lockdown. The general weekly drops have also been observed for the truck flow on the highways by 1.63% the first week, then it raises back to 18.31% the second week, and declined again by 10.98% the third week, and declined by 2.55% at the fourth week following the county-wide lockdown. Accordingly, the change of traffic trigged the intensive decline of $NO_2$ by 44.23% the first week, 12.96% the second week, 50.64% the third week, 32.65% the fourth week following the lockdown; We found a dramatic drop in $PM_{2.5}$, $NO_2$, CO during the **first** week after initiating the stay at home order. The correlation (Pierson r) between truck flow change and changes of $NO_2$, CO, and $PM_{2.5}$ is 0.91(****),0.88(****),0.74(**); The correlation between traffic flow change and changes of $NO_2$ is 0.87(****), CO is 0.81(***), and $PM_{2.5}$ is 0.62(**). The correlation between traffic speed change and changes of $NO_2$ is -0.84(****), CO is -0.78 (***), and $PM_{2.5}$ is -0.59(*). We found that a decline of 1% in $NO_2$, CO and $PM_{2.5}$ is associated with the decline of 15.79%, 17.15% and 9.43% in truck flow; A decline of 1% in $NO_2$, CO and $PM_{2.5}$ is associated with the decline of 11.26%, 9.43% and 20.96% in traffic flow;

A decline of 1% in $NO_2$, CO and $PM_{2.5}$ is associated with the increase of 3.45%, 3.13% and 5.21% in traffic speed. The results are all statistically significant.

**Conclusion**: The drop of truck flow is mainly responsible for the drop of $NO_2$ and CO. The lockdowns provided a large-scale experiment into air quality research. The result of this research would provide an important reference for the policy markers regarding truck management in light of air quality control to prepare the 2028 Summer Olympics in LA.

## 1. Introduction

The sudden outbreak of COVID-19 on the global scale has shown a glimpse of a cleaner world as indicated in many reports. However, visual perception alone can be deceptive when observing ambient air quality (IQAir,2020).While the perception of a better air quality might be more obvious in developing countries such as China and India, high quality data with fine resolution need to be studied to quantify the anthropogenic impacts on the air quality research.

It has long been known that vehicular emissions degrade air quality, which has lingering health effects (Zhang et al., 2010). Nitrogen Dioxide ($NO_2$) and Carbon monoxide (CO) reduces oxygen transport in the bloodstream when inhaled, and is emitted from combustion processes, with vehicles representing a huge contribution in urban domains like Los Angeles (LA) (Marshall et al., 2003). Particulate matter with aerodynamic diameter less than 2.5 μm ($PM_{2.5}$) has been linked with cardiovascular and respiratory disease and can decrease visibility and cause material damage (Feenstra et al.,2019). Understanding the natural and anthropogenic emission sources of $NO_2$, CO and $PM_{2.5}$ emissions in urban regions are important for assessing exposure and designing mitigation strategies (Newman et al., 2013). A recent study by Harvard university indicated that a small increase in the long-term exposure to $PM_{2.5}$ leads to a large increase in the COVID-19 death rate. (Wu et al., 2020)

The LA basin only covers ~4% of land area of California, it encompasses more than 43% of the state's population (Wong et al., 2015). The basin is surrounded on three sides by mountains and bounded by the Pacific Ocean (Wunch et al., 2009). Some of the most polluted air in USA is contained in the South Coast air basin (SCB) (Wunch et al., 2009). The long-term exposure to $PM_{2.5}$ may not only adversely affect the respiratory and cardiovascular system and increase mortality risk; it also exacerbates the severity of COVID-19 infection symptoms. (Wu et al., 2020) Satellite column density data are unable capture the hourly spatial heterogeneity of regional pollutants from traffic. Due to a lack of hourly spatial coverage from traditional air monitors, most studies use models to estimate traffic emission rather than traffic proximity measures (Wilhelm et al., 2011b). However, some studies showed that while simulation models performed reasonably well for $NO_2$, CO, $PM_{2.5}$, their emissions were underestimated (Zhang et al.,2010).

On March 26, 2020 the US EPA announced that power plants, factories and other facilities are now allowed to determine for themselves if they are able to meet legal requirement for air quality. (Wu et al., 2020). The lockdown generally resulted in much lower emission of pollutants related to mobile sources, but other sources of air pollution have remained stable or even increased. Therefore, it is not only important to quantify the change in pollutants due to the lockdowns but also to track the change of $PM_{2.5,}$ which could have shared an important linkage to the death rate of the coronavirus.

In this study, we hypothesize that because of the lockdown, the emissions of $NO_X$ and CO have significantly dropped, which might lead to the increasement of $O_3$. The lockdown would also decrease the $PM_{2.5}$ concentration because of the decline of daily commuting. However, because of the increase of household electricity consumption, the $PM_{2.5}$ could stay stable or even increase during some period of time of COVID-19 lockdown. We determine the spatial structure of $NO_2$, CO, $O_3$ and $PM_{2.5}$ using a monitoring network from South Coast AQMD. We pinpoint the locations of high pollution exposure, and assess their diurnal and nightly variability during the outbreak of COVID-19. This type of analysis allows for neighborhood scale planning and health risk assessment. The lockdowns provided a large-scale experiment into air quality research. The result of this research would provide an important reference for the policy markers regarding to traffic management in light of environmental health benefits.

## 2. Method

### 2.1 The California Ambient Air Monitoring Network

The California Ambient Air Monitoring Network consists of more than 250 monitoring stations operated by federal, State, and local agencies. The network has been dedicated to measuring ambient concentrations of criteria pollutants including ground-level ozone($O_3$), particulate matter ($PM_{10}$ and $PM_{2.5}$), carbon monoxide (CO), nitrogen dioxide ($NO_2$), sulfur dioxide ($SO_2$) and lead (Pb). (California Air Resource Board, 2019) The distribution of the California Ambient Air Monitoring Network in LA Basin is in Figure1. The data is currently operated by South Cast AQMD and can be downloaded at: http://www.aqmd.gov/aq-spec.

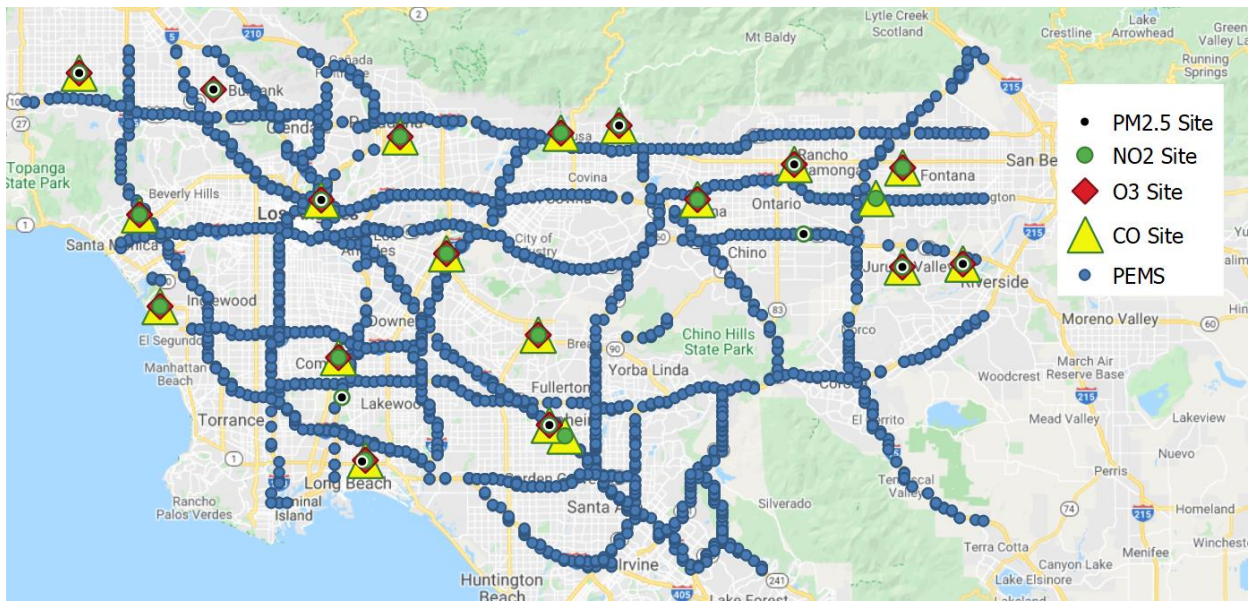

**Figure 1. The distribution of sensors in LA Basin. The yellow triangle represents CO targets; The blue dots represents 3144 PeMS sensors on freeway; The green dots represent $NO_2$ targets; The red rhombus represents the $O_3$ site; The tiny black dots represent $PM_{2.5}$ site.**

### 2.2 PeMS Data

In order to validate the measurement variability of $NO_2$, CO, $O_3$ and $PM_{2.5}$ associated with emission source, the real-time based freeway traffic dataset from the California Department of Transportation (PeMS, http://pems.dot.ca.gov ) is obtained. PeMS collects data from varies types of vehicle detector stations, including inductive loops, side-fire radar and magnetometers. (Caltrans, 2013.) PeMS can provide flow and speed as reported by detectors over several years. It also supports integration with common internet-based mapping service (e.g., Google Maps, Google Earth). (Caltrans, 2013.)In this research, a python-based script is created to fetch the data of PeMS collocated with freeways around LA basin. There are around 3144 sensors in this area. The distribution of PeMS sites within LA basin is shown in Figure1. In this study, the hourly traffic speed, traffic flow and truck flow data from January to April in both 2019 and 2020 has been used to study the change of traffic pattern in hourly, daily and monthly pattern. It has been pointed out in a previous study that when examining the %VMT-speed distribution for the entire freeway network in Los Angeles County for an entire average day, it is found that speeds around 65 to 70 mph dominate. This implies that the freeway system is operating in a good condition. (Barth and Boriboonsomsin, n.d.)

## 2.3 Data Processing

We used hourly data for all sensors in LA basin to calculate and compare the daily average difference of the traffic flow, truck flow, traffic speed, $NO_2$, $CO$ $O_3$ and $PM_{2.5}$ during first 4 months of 2019 and 2020 (Fig 2 and Fig 3). The variance for the timeseries fig came from time and spatial heterogeneity.

Then we calculated the monthly trend of those 4 variables in box plot in order to get a general sense of data in the monthly means (Fig 4 and Fig 5). The variance for the timeseries fig came from time and spatial heterogeneity.

Since the traffic pattern is highly influenced by weekday and weekend pattern, therefore the weekly difference percentage ratio between 2019 and 2020 is calculated to better understand the impact of COVID-19 on different pollutants. (Fig 6) The reason to compare with the weekly trend instead of the daily trend is that the traffic and traffic driven- pollutants are weekly dominated. On the daily scale, it is likely the daily difference came from the difference between weekdays and weekends. In order to offset the alternative influence of weekdays and weekends, therefore the weekly trends are compared. Since here we average the value in weekly format in order to offset the timely difference between weekdays and weekends, therefore the variance of the weekly difference percentage ratio between 2019 and 2020 came from the spatial heterogeneity only. The difference percentage ratio between 2019 and 2020 for a random variable in a specific sensor can be calculated as:

$$y_{n_i} = \left(\frac{x_{n_{i_{2020}}}}{x_{n_{i_{2019}}}} - 1\right) \times 100\% \qquad (2.3.1)$$

Here, $y_{n_i}$ represents the difference percentage ratio in week $i$ of 2020 comparing with the week $i$ of 2019 for sensor $n$ for a random variable; $x_{n_{i_{2020}}}$ is a random variable value for the sensor $n$ in the week $i$ of 2020, $x_{n_{i_{2019}}}$ is the variable value for the sensor $n$ in week $i$ of 2020.

In order the calculate the ratio mean and variance for all sensors for a specific week. Here, we followed the method in Kendall's Advanced Theory of Statistics (Stuart et al.,1998 ) and Survival Models and Data Analysis (Elandt-Johnson and Johnson, 1980). Given random variables in week i of 2020 and 2019: $X_{i_{2020}}$ and $X_{i_{2019}}$ where $X_{2019}$ either has no mass at 0 (discrete) or has support $[0, \infty)$. Let $G = g(X_{i_{2020}}, X_{i_{2019}})$. The approximations for the mean of the ratio of random variable $X$ in week $i$ of 2020 and 2019 $E(X_{i_{2020}}/X_{i_{2019}})$ after 2 Taylor expansion can be improved approximately as:

$$E(X_{i_{2020}}/X_{i_{2019}}) \approx \frac{\mu_{x_{i_{2020}}}}{\mu_{x_{i_{2019}}}} - \frac{Cov(X_{i_{2020}}, X_{i_{2019}})}{(\mu_{x_{i_{2019}}})^2} + \frac{Var(\mu_{x_{i_{2019}}})\mu_{x_{i_{2020}}}}{(\mu_{x_{i_{2019}}})^3} \qquad (2.3.2)$$

$\mu_{x_{i_{2020}}}$ and $\mu_{x_{i_{2019}}}$ are the mean the random variable $X$ in week $i$ of 2020 and 2019; $Cov(X_{i_{2020}}, X_{i_{2019}})$ is the covariance of random variable $X$ in week $i$ of 2020 and 2019; $Var(\mu_{x_{i_{2019}}})$ is the variance of random variable $X$ in week $i$ of 2019.

The approximations of the variance of the ratio of random variable $X$ in week $i$ of 2020 and 2019 $Var(X_{i_{2020}}/X_{i_{2019}})$ after 2 Taylor expansion can be improved approximately as:

$$Var(X_{i_{2020}}/X_{i_{2019}}) \approx \frac{(\mu_{x_{i_{2020}}})^2}{(\mu_{x_{i_{2019}}})^2}\left[\frac{Var(\mu_{x_{i_{2020}}})}{(\mu_{x_{i_{2020}}})^2} - 2\frac{Cov(X_{i_{2020}}, X_{i_{2019}})}{\mu_{x_{i_{2020}}}\mu_{x_{i_{2019}}}} + \frac{Var(\mu_{x_{i_{2019}}})}{(\mu_{x_{i_{2019}}})^2}\right] \quad (2.3.3)$$

$Var(\mu_{x_{i_{2019}}})$ is the variance of random variable $X$ in week $i$ of 2020. According to equation 2.3.1, in week $i$ the average difference percentage ratio between 2019 and 2020 for all sensors can be summarized as:

$$E(X_{i_{2020}}/X_{i_{2019}} - 1) \approx \left[\frac{\mu_{x_{i_{2020}}}}{\mu_{x_{i_{2019}}}} - \frac{Cov(X_{i_{2020}}, X_{i_{2019}})}{(\mu_{x_{i_{2019}}})^2} + \frac{Var(\mu_{x_{i_{2019}}})\mu_{x_{i_{2020}}}}{(\mu_{x_{i_{2019}}})^3} - 1\right] \times 100\% \quad (2.3.4)$$

In week i the variance of difference percentage ratio between 2019 and 2020 for all sensors can be summarized as:

$$Var(X_{i_{2020}}/X_{i_{2019}} - 1) \approx \frac{(\mu_{x_{i_{2020}}})^2}{(\mu_{x_{i_{2019}}})^2}\left[\frac{Var(\mu_{x_{i_{2020}}})}{(\mu_{x_{i_{2020}}})^2} - 2\frac{Cov(X_{i_{2020}}, X_{i_{2019}})}{\mu_{x_{i_{2020}}}\mu_{x_{i_{2019}}}} + \frac{Var(\mu_{x_{i_{2019}}})}{(\mu_{x_{i_{2019}}})^2}\right] \times 10\% \quad (2.3.5)$$

The calculation result for 2.3.4 and 2.3.5 can be found at Table 1. Following after Table 1, we calculated the correlation between each 2 variables in fig 7.

Additionally, heatmaps in 24 hourly and monthly mean patterns are also utilized here to understand the geospatial changes during the COVID-19 outbreak. Finally, the correlation among different variables are made to determine the extent of traffic impacts on different pollutants. The time for the data used in this study has been adjusted to account for varying day length.

## 3. Results

In Fig 2, we compared the traffic pattern average daily variability between the year of 2019 and the year of 2020. Both traffic and truck flow showed strong week and weekend pattern during 2019 and 2020. The average number of vehicles per hour in the LA basin has been stabilized approximately between 3000 to 4000 while the truck flow has been stabilized approximately between 100 to 210 until March 14, 2020 Saturday (Day of Year: 74), which was 4 days before the stay at order. Starting from March 15, 2020 the traffic flow has dropped significantly from the year of 2019 on a weekly basis. The similar trend has been also observed in truck flow. Conversely, traffic speed has maintained roughly same as the year of 2019 until March 14, 2020 Saturday (Day of Year: 74). Starting from March 15, 2020 the traffic speed has been enhanced compared with last year. The variance indicated that there is a significant time and spatial heterogeneity for traffic flow, truck flow and traffic speed. This heterogeneity can be also observed by the spatial distribution map in Fig 10 and Fig 11.

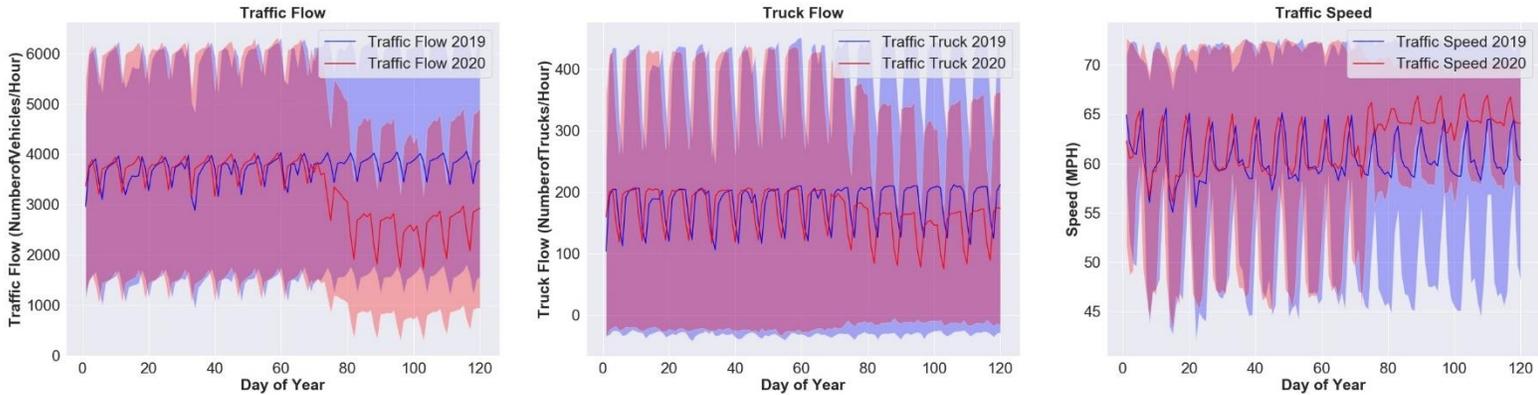

**Fig 2. Daily traffic pattern between 2019 to 2020, from left to right are: traffic flow, truck flow and traffic speed, the x axis represents the number of days from January 1. The unit for both traffic and truck flow is number of vehicles (trucks) per hour; The unit for traffic speed is miles per hour. The shadow of three plots represent the standard deviation.**

In Fig 3, we compare the day-to-day variability in daily average pollutant concentrations between 2019 and 2020. As can be seen from Figure 3.1, the ozone trend has been positive from Jan to April as is expected, and there is not much difference in ozone between 2019 and 2020 and $NO_2$ and CO generally showed a negative trend from Jan to Apr. Before Saturday March 14, 2020 (Day of Year: 74) the $NO_2$ and CO concentration in 2020 were generally larger than the same period in 2019. Starting from March 15, 2020 the $NO_2$ and CO concentration in 2020 were generally less than during the same period in 2019. **Therefore, the shifts in $NO_2$ and CO trends share the same timing due to the impact of the lockdown.** The trend for $PM_{2.5}$ was generally higher during first 2 months of 2019 and 2020 because of the heating usage during the winter. Before March 7, 2020 Saturday (Day of Year: 67) the $PM_{2.5}$ concentration in 2020 was generally larger than the same period in 2019. Starting from March 8, 2020 the $PM_{2.5}$ concentration in 2020 is generally less than the same period in 2019. The timing of the shift for $PM_{2.5}$ behavior during 2020 relative to 2019 was one week ahead of that for $NO_2$, CO and traffic, potentially indicating the importance of a different source for $PM_{2.5}$ than for the other pollutants.

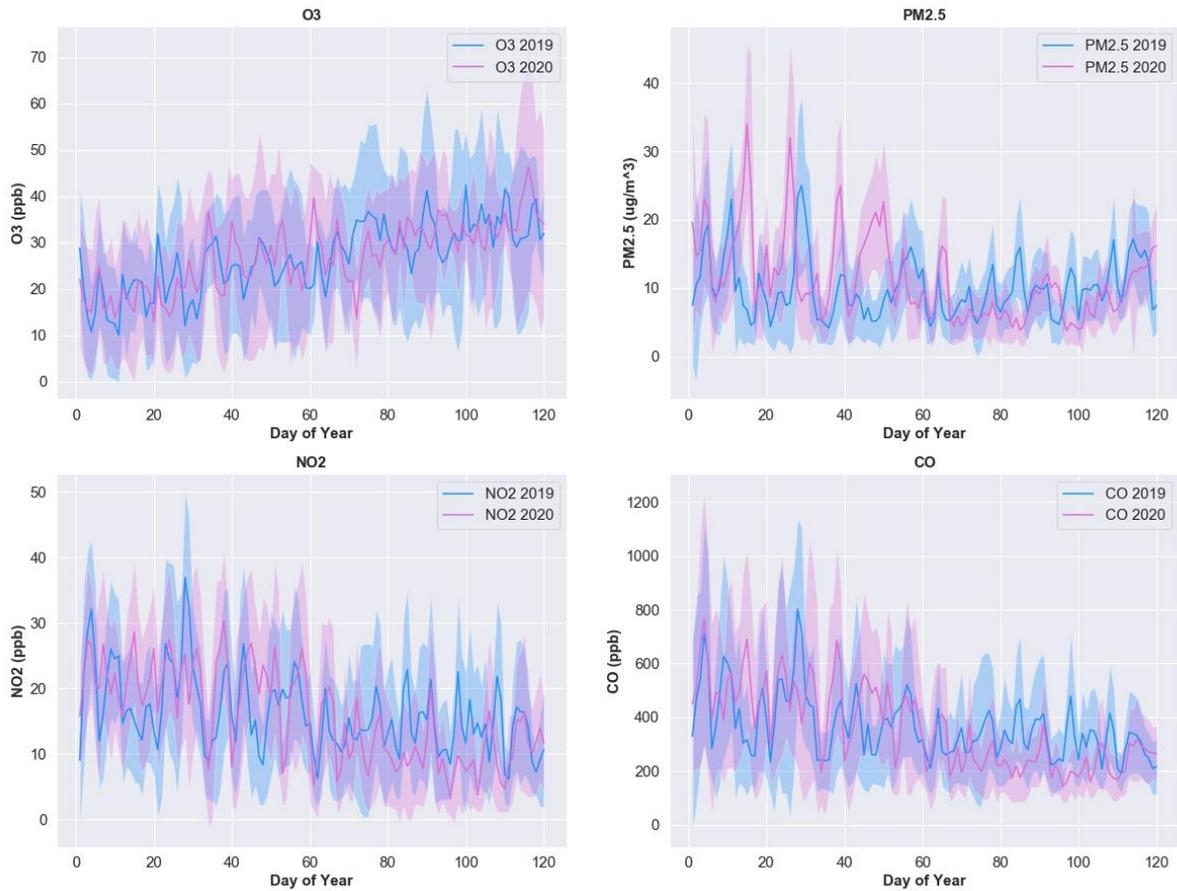

**Fig 3. Daily traffic pattern between 2019 to 2020, from left to right are: traffic flow, truck flow and traffic speed, the x axis represents the number of days from January 1. The unit for both traffic and truck flow is number of vehicles(trucks) per hour; The unit for traffic speed is miles per hour. The shadows of plots represent the standard deviation σ.**

Figures 4 and 5 reflect the monthly distribution for 7 variables in box plot format. In the box plots, the median is shown by the line dividing the box into two parts. Values outside the interquartile range (indicated by the boxes) are shown as whiskers. In Fig 4, the traffic flow plot is relatively tall suggests the traffic situation varies with location and by day. The long interquartile range in traffic flow, truck flow and traffic speed indicate that there are some traffic hotspots in specific locations. Those difference would be elaborated in the discussion part. In figure 4 and 5, there is also a difference between 2019 and 2020 during the March and April box plots in terms of traffic flow, traffic speed, $PM_{2.5}$, $NO_2$, which indicated there is unusual difference between those two groups due to the lockdown. The $O_3$ plots between 2020 and 2019 have similar size and median and both are evenly distributed indicated the difference between those 2 groups are similar. During Jan and Feb, all figures have similar median, but are slightly different in terms of distribution, which means further investigation need to be initiated to understand the trend at the micro level.

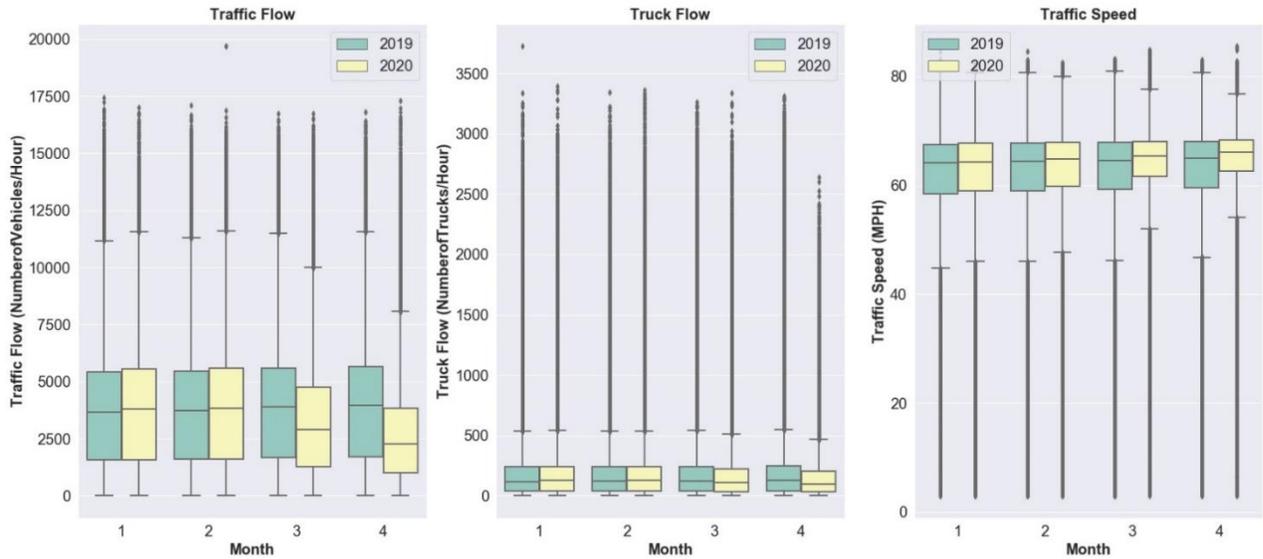

**Figure 4. Box plots for Traffic Flow, Truck Flow and Traffic Speed from Jan to April in 2019 and 2020.**

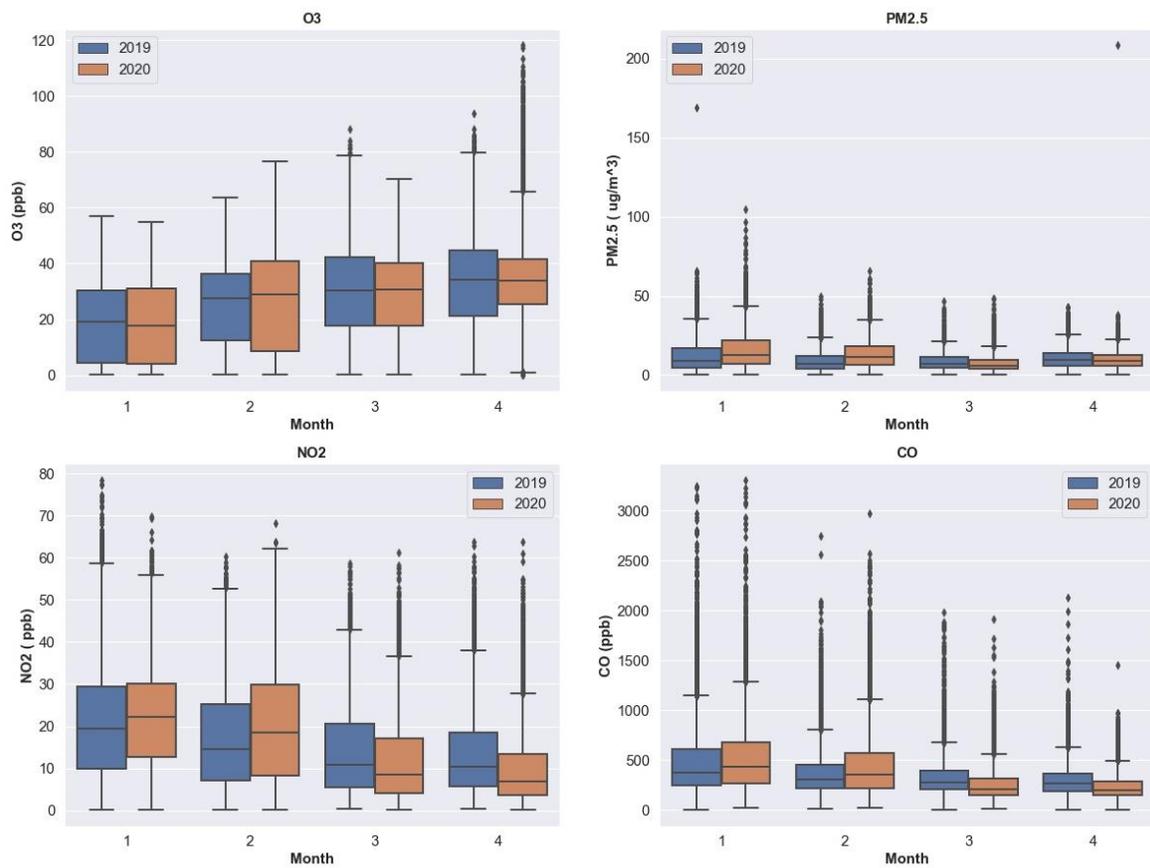

**Figure 5. Box plots for O$_3$, PM$_{2.5}$, NO$_2$ and CO from Jan to April in 2019 and 2020.**

Fig 6 shows the average weekly percentage change for each sensor of for traffic flow, truck flow, traffic speed, O$_3$, PM$_{2.5}$, NO$_2$ and CO in from the second to 17th week of 2020 compared with 2019. The way to calculate the average weekly percentage change for each sensor in equation 2.3.4 reflects the more details in spatial heterogeneity. The x axis started from the second week of 2020 since the first week did not contain a full 7-day in 2020. As the average difference ratio of traffic flow for each sensor started dropping intensely in week 10, the general trends for $PM_{2.5}$, $NO_2$ and CO have been trigged to drop week by week below 0. The average truck flow for each sensor is generally higher in 2020 than 2019 and the trend of difference ratio started dropping in week 10. As the trend for truck flow resumed to another peak on week14, the general trends for $PM_{2.5}$, $NO_2$ and CO have been trigged to resume back to peak on week14. The trend for $O_3$ is normally opposite to that for $NO_2$ and CO.

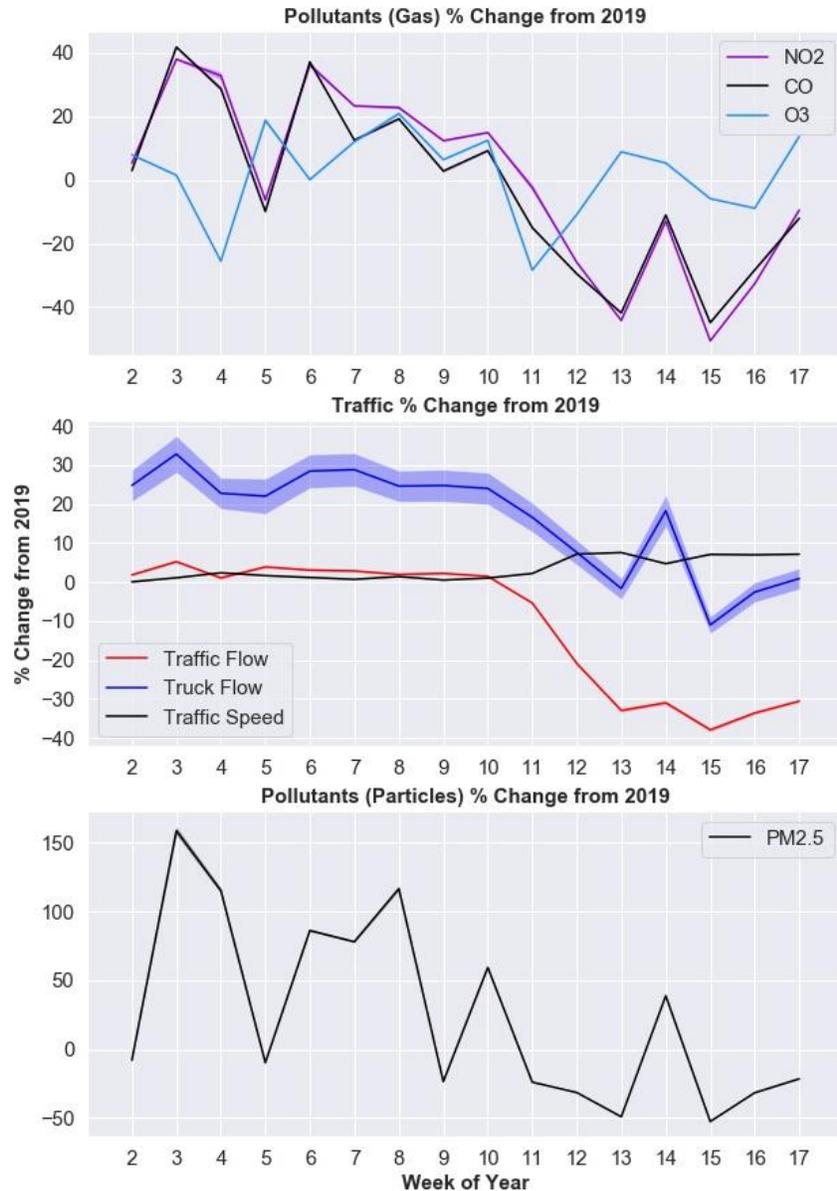

**Figure 6. The average weekly percentage change for traffic flow, truck flow, traffic speed, $O_3$, $PM_{2.5}$, $NO_2$ and CO for each sensor from the 2nd to the 17th week of 2020 compared with the same week in 2019. The shadows of plots represent the variance $\sigma^2$.**

According to Table 1, the traffic flow on highways started to drop intensely by 20.86 % when initiating the stay at home order and it continued decreasing by 32.92% the first week, 30.94% the second week, 37.9% the third week, and 33.57% at the fourth week following the lockdown compared to corresponding dates during the year of 2019. The average truck flow for each sensor is generally higher in 2020 than 2019 before lockdown. The general weekly drops have also been observed for the truck flow on the highways by 1.63% the first week, then it raises back to 18.31% the second week, and declined again by 10.98% the third week, and declined by 2.55% at the fourth week following the county-wide lockdown. Accordingly, the change of traffic trigged the intensive decline of $NO_2$ by 44.23% the first week, 12.96% the second week, 50.64% the third week, 32.65% the fourth week following the lockdown; CO declined by 41.85% the first week, 11.04% the second week, 44.92% the third week, 28.23% the fourth week following the lockdown; $PM_{2.5}$ declined by 48.98% the first week, raises back to 38.84% the second week, and declined again by 52.43% the third week, and declined by 31.55% the fourth week following the lockdown;

**Table 1 The Weekly Percentage Change for Different Variables in 2020 Compared to 2019**

| Week of 2020 | Traffic Flow Change | | Speed Change | | Truck Flow Change | | $PM_{2.5}$ Change | | $O_3$ Change | | $NO_2$ Change | | CO Change | |
|---|---|---|---|---|---|---|---|---|---|---|---|---|---|---|
| | Mean | $\sigma^2$ | Mean | $\sigma^2$ | Mean | $\sigma^2$ | Mean | $\sigma^2$ | Mean | $\sigma^2$ | Mean | $\sigma^2$ | Mean | $\sigma^2$ |
| 2 | 1.85% | 0.23% | 0.07% | 0.03% | 24.80% | 4.05% | -7.98% | 0.20% | 7.94% | 0.33% | 5.29% | 0.19% | 2.92% | 0.30% |
| 3 | 5.25% | 0.25% | 1.11% | 0.03% | 32.85% | 4.69% | 158.60% | 3.40% | 1.49% | 0.29% | 38.03% | 0.33% | 41.83% | 0.50% |
| 4 | 1.08% | 0.23% | 2.38% | 0.03% | 22.79% | 3.97% | 115.18% | 2.23% | -25.56% | 0.22% | 32.83% | 1.33% | 28.73% | 0.93% |
| 5 | 3.89% | 0.28% | 1.67% | 0.03% | 22.06% | 4.51% | -9.73% | 0.08% | 18.77% | 0.83% | -6.33% | 0.40% | -9.91% | 0.53% |
| 6 | 3.09% | 0.24% | 1.18% | 0.02% | 28.48% | 4.31% | 86.20% | 0.87% | 0.10% | 0.25% | 36.23% | 0.93% | 37.15% | 0.68% |
| 7 | 2.86% | 0.24% | 0.72% | 0.02% | 28.84% | 4.26% | 78.06% | 0.90% | 12.03% | 0.28% | 23.33% | 0.36% | 12.54% | 0.29% |
| 8 | 1.97% | 0.24% | 1.42% | 0.02% | 24.63% | 3.98% | 116.59% | 1.67% | 20.89% | 0.42% | 22.80% | 0.51% | 19.24% | 0.42% |
| 9 | 2.25% | 0.23% | 0.53% | 0.02% | 24.78% | 4.09% | -23.39% | 0.37% | 6.39% | 0.20% | 12.38% | 0.42% | 2.82% | 0.55% |
| 10 | 1.48% | 0.23% | 1.04% | 0.02% | 24.04% | 4.08% | 59.38% | 0.62% | 12.48% | 0.12% | 14.92% | 0.26% | 9.23% | 0.35% |
| 11 | -5.36% | 0.23% | 2.22% | 0.03% | 16.62% | 3.74% | -23.85% | 0.21% | -28.40% | 0.07% | -2.39% | 1.16% | -15.01% | 0.49% |
| 12* | -20.86% | 0.30% | 7.18% | 0.07% | 7.63% | 3.19% | -31.39% | 0.11% | -10.87% | 0.08% | -25.94% | 0.18% | -29.55% | 0.14% |
| 13 | -32.92% | 0.41% | 7.58% | 0.08% | -1.63% | 2.73% | -48.98% | 0.08% | 8.87% | 0.14% | -44.23% | 0.21% | -41.85% | 0.11% |
| 14 | -30.94% | 0.46% | 4.73% | 0.06% | 18.31% | 4.09% | 38.84% | 0.21% | 5.34% | 0.18% | -12.96% | 0.14% | -11.04% | 0.23% |
| 15 | -37.90% | 0.42% | 7.11% | 0.07% | -10.98% | 2.20% | -52.43% | 0.16% | -5.89% | 0.11% | -50.64% | 0.11% | -44.92% | 0.14% |
| 16 | -33.57% | 0.35% | 7.01% | 0.06% | -2.55% | 2.55% | -31.55% | 0.08% | -8.88% | 0.08% | -32.65% | 0.17% | -28.32% | 0.17% |
| 17 | -30.53% | 0.31% | 7.16% | 0.06% | 0.92% | 2.66% | -21.42% | 0.38% | 13.68% | 0.30% | -9.49% | 0.33% | -12.06% | 0.34% |

\* and color red represents the week initiated and after stay at home order from the state government of California

In figure 7, we did a regression analysis to identify the correlation between different variables. The correlation (Pierson r) between truck flow change and changes of $NO_2$, CO, $PM_{2.5}$ and $O_3$ is **0.91(\*\*\*\*)**,0.88(\*\*\*\*),0.74(\*\*),0.2(ns); The correlation between traffic flow change and changes of $NO_2$ is 0.87(\*\*\*\*), CO is 0.81(\*\*\*), $PM_{2.5}$ is 0.62(\*\*)and $O_3$ is 0.14 (ns); The correlation between traffic speed change and changes of $NO_2$ is -0.84(\*\*\*\*), CO is -0.78 (\*\*\*), $PM_{2.5}$ is -0.59(\*)and $O_3$ is -0.19(ns). Therefore, the traffic has the most impact on $NO_2$, followed by CO, followed by $PM_{2.5}$. The correlation between truck flow and $NO_2$ is most statistically significant. Assuming the relation between $NO_2$, CO, $PM_{2.5}$ and truck flow is linear, according to the regression line in figure 7, we can conclude that a decline of 1% in $NO_2$, CO and $PM_{2.5}$ is associated with the decline of 15.79%, 17.15% and 9.43% in truck flow; A decline of 1% in $NO_2$, CO and $PM_{2.5}$ is associated with the decline of 11.26%, 9.43% and 20.96% in traffic flow; A decline of 1% in $NO_2$, CO and $PM_{2.5}$ is associated with the increase of 3.45%, 3.13% and 5.21% in traffic speed.

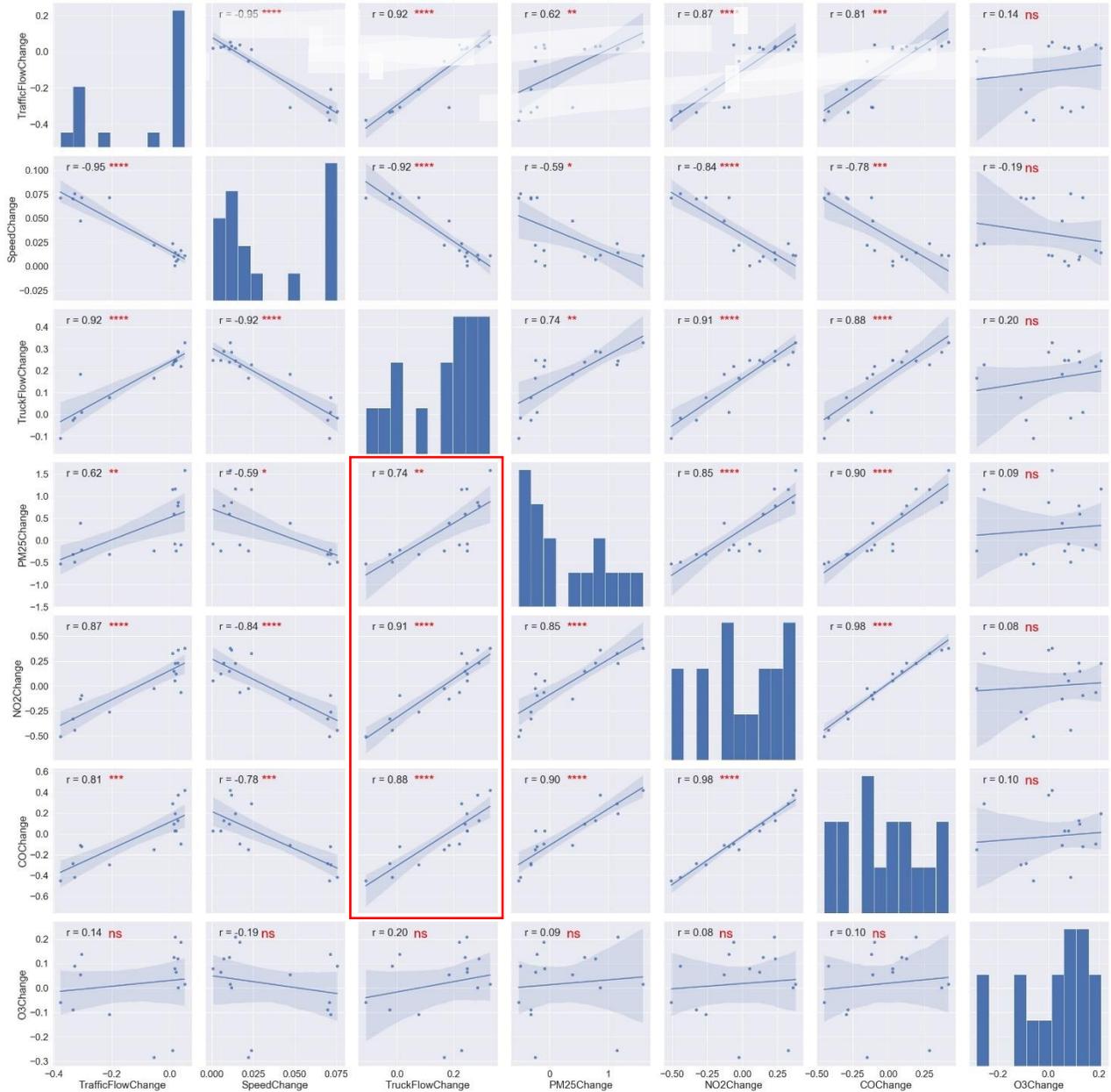

**Fig 7. Regression analysis for the percentage change of traffic flow, truck flow, traffic speed, $O_3$, $PM_{2.5}$, $NO_2$ and CO in from the second to 17th week of 2020 comparing with the year of 2019. The shadow of the regression line represents the 95% confidence intervals (95% CI). * and ns indicated P value. (ns: P>0.05; *: P ≤ 0.05; **:P ≤ 0.01; ***:P≤ 0.001; **** P≤ 0.0001)**

## 4. Conclusion

At the macro level, the $O_3$ trend is quite different from that of the other pollutants. From January to April 2020, $O_3$ increases in a similar manner to 2019. This suggests that ozone during this time of year is controlled by sources other than by traffic emissions. At the micro level, $PM_{2.5}$ has multiple drivers in different seasons besides traffic. Due to demand for heating during the winter, $PM_{2.5}$ is generally higher in January and February than in April and March. After initiating the state-level lockdown, $PM_{2.5}$ dropped: However, this decline changed to a rise again during the stay at home order. This might have been due to the increased usage of home utilities. $NO_2$ and CO have been clearly been most affected by truck and traffic flow. Our study also quantifies the tie between $NO_2$ and traffic flow by indicating that a decline of 1% in $NO_2$, CO and $PM_{2.5}$ is associated with a decline of 15.79%, 17.15% and 9.43% in truck flow; A decline of 1% in $NO_2$, CO and $PM_{2.5}$ is associated with the decline of 11.26%, 9.43% and 20.96% in traffic flow; A decline of 1% in $NO_2$, CO and $PM_{2.5}$ is associated with an increase of 3.45%, 3.13% and 5.21%% in traffic speed in LA Basin, which could be treated as an important reference for policy marker to execute traffic control for the environmental benefit. During the COVID-19 lockdown period, we saw the rising trend of truck flow ratio difference in the week 14 (Mar 30,2020 to Apr 5, 2020) which triggered the rising trend of $PM_{2.5}$, $NO_2$, and CO. This is mainly due to the arrival of hospital ship Mercy, which may potentially increase the truck flow to help with the establishment the hospital infrastructure and delivery of the emergency facilities.

As the previous study indicated the potential linkage between $PM_{2.5}$ and mortality rate of COVID-19, the amount of traffic flow should be controlled under a certain level for the protection of public health, which would be another important reference for the policy in this specific period.

## 5. Discussion

As mentioned in the Results Section regarding Figures 5 and 6, the traffic data (especially truck) have a larger range than others because there are more sensors to reflect the spatial heterogeneity. Also, since all figures have similar median, but are slightly different in terms of distribution, which means further investigation needs to be initiated to understand the trend at the micro level. Therefore, we are using this supplement (?) to discuss the hourly and spatial heterogeneity for those 7 variables.

For the 24-hour patterns in Figures 8 and 9, each column has 4 panels from left to right represent January to April. Each panel has been evenly divided into 24 slices; each slice represents an hour in clockwise (upper right: 0:00-5:00, down right: 6:00-11:00, down left: 12:00-17:00, upper left:18:00-23:00. Each panel consists of 29-31 circles. Every circle in each panel represents a specific date in this month(panel). From inside to out, the first inner circle with the least diameter represents the first day of a specific month, the last outside circle with the largest diameter represents the last day of a specific month. For example, the first inner circle in the first row and first column represents traffic flow value of Jan 01, 2020 Wednesday; Likewise, the last outside circle with the largest diameter represents traffic flow value of Friday Jan 31, 2020.

In Figure 8, from top to bottom, each row represents the value of traffic flow, truck flow and traffic speed. In January and February 2020, the morning commute starts around 5:00 and reaches the morning peak at around 7:00; After the morning commute, the truck flow reached to peak between 11:00-12:00. As the commute continued, the afternoon commute reaches a second peak between 15:00-17:00. Then the traffic situation relieved after 19:00. The high traffic congestions with low traffic speed occur during peak hours. The weekday-weekend pattern was very obvious in the first 2 months of 2020 too. Fridays and weekends can be easily identified with low value (blue) in regular peak hours alternatively every 6-8 days. In March to April, as the city lockdown, the high traffic and truck values and low traffic speed during peak hours gradually faded away week by week after the announcement of the stay at home order. However, the week-weekend cycle still exists as some essential people such as delivery people still maintain the commuting behaviors.

In Figure 8, from top to bottom, each row represents the value of $O_3$, $NO_2$, CO and $PM_{2.5}$. The $O_3$ starts accumulating after sunrise between 7:00 – 8:00, it usually increases until 14:00. From January to April, due to the increasing daylength, high values of ozone extended from 9:00-18:00 to 8:00-20:00.

The $NO_2$ and CO started accumulating during the morning commute. Also because of the effects of the lockdown, there is a significant drop for $NO_2$ and CO during March and April.

$PM_{2.5}$ is generally higher in the Jan and Feb because of heating and it declines in March and April mainly because of the county-wide shutdown. However, by the end of April, the $PM_{2.5}$ increased, perhaps due to the slight increase of traffic flow, to which the policy markers should pay attention, since there is a potential linkage between $PM_{2.5}$ and mortality rate of coronavirus.

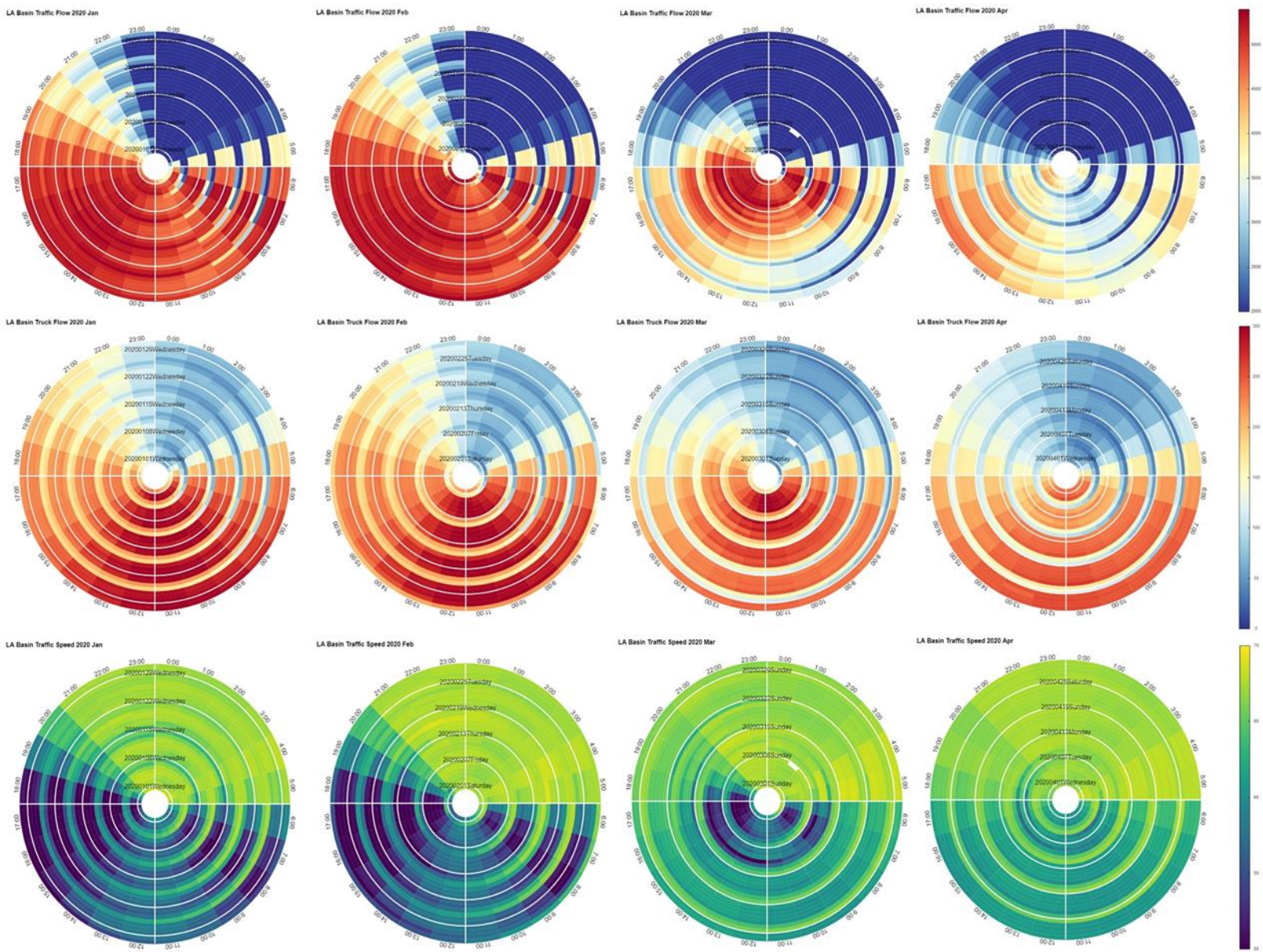

**Fig 8. 24-hour pattern for Traffic Flow, Truck Flow and Traffic Speed from Jan to April in 2020**

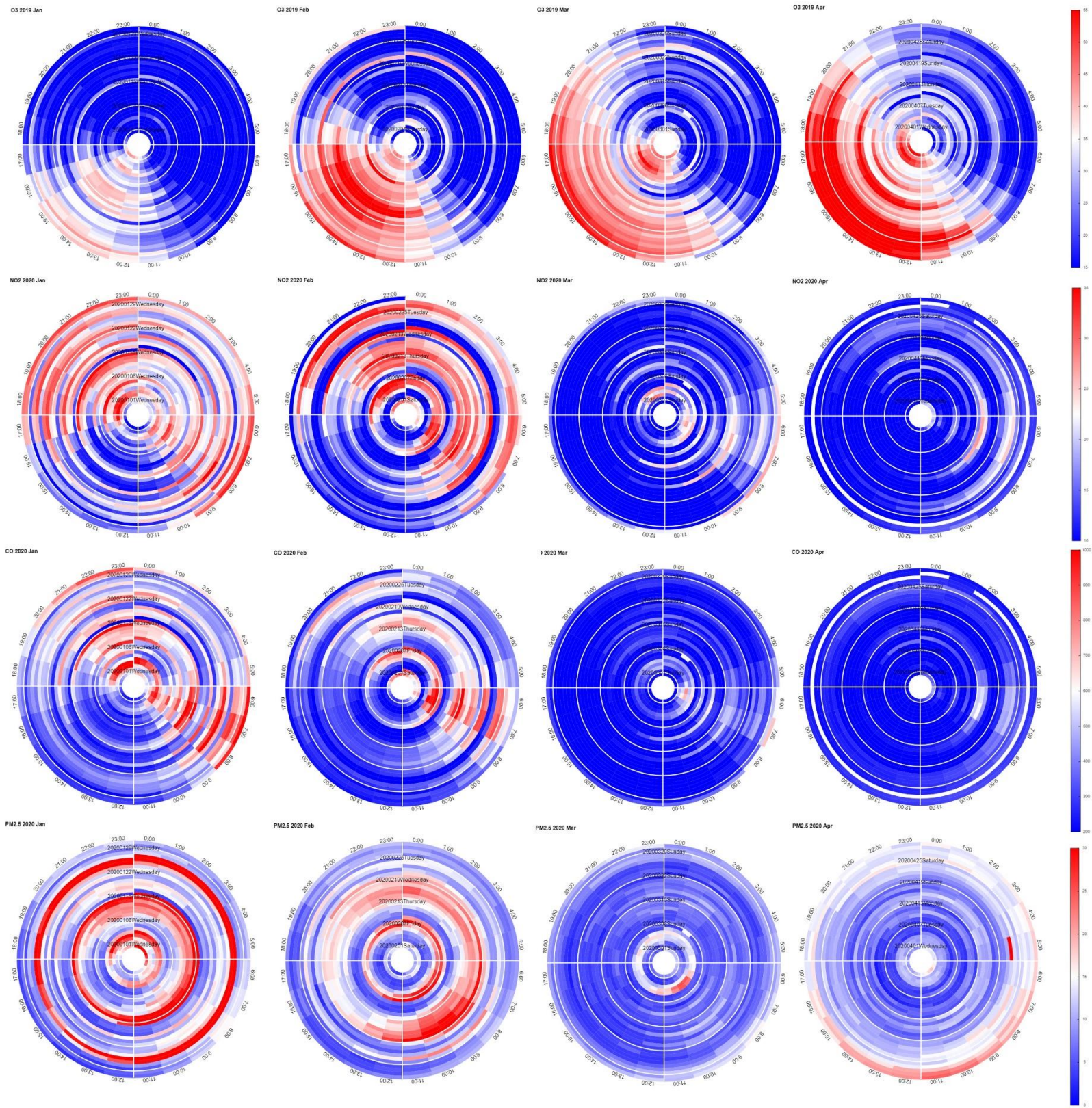

**Fig 9. 24-hour pattern for $O_3$, $NO_2$, CO and $PM_{2.5}$ from Jan to April in 2020**

Figures 10 and 11 show spatial heterogeneity for those 7 variables. In Figure 10, from top to bottom each row represents the value of traffic flow, truck flow and traffic speed. Before lockdown, the high traffic congestion areas were mainly concentrated by the coast and LA downtown. After the announcement of the stay at home order, traffic hotspots gradually vanished month by month. In April, hotspots of traffic were mainly concentrated in the downtown area and Ontario where the most warehouses are concentrated.

As showed in fig 10, truck flow has larger spatial heterogeneity than traffic flow and traffic speed. This explained well the long interquartile for truck flow in fig 4 indicated hot spots in fig 10. When looking at LA as a whole point from fig 2, it seems the truck flow has declined from 2019 before lockdown. However, when comparing the change ratio from each PeMS sensor, it tells a different story in Table 1: The average change ratio of truck flow from 2019 for each sensor actually increased. This is due to the large spatial heterogeneity for truck flow showed in fig 10.

In Figure 10 from top to bottom, each row represents the value of $O_3$, $NO_2$, CO and $PM_{2.5}$. The basic trend is similar as discussed for the 24-hour trend as well as results. Some hotspots can be identified related to the commute and truck hotspots, which are mainly concentrated in downtown and warehouse districts.

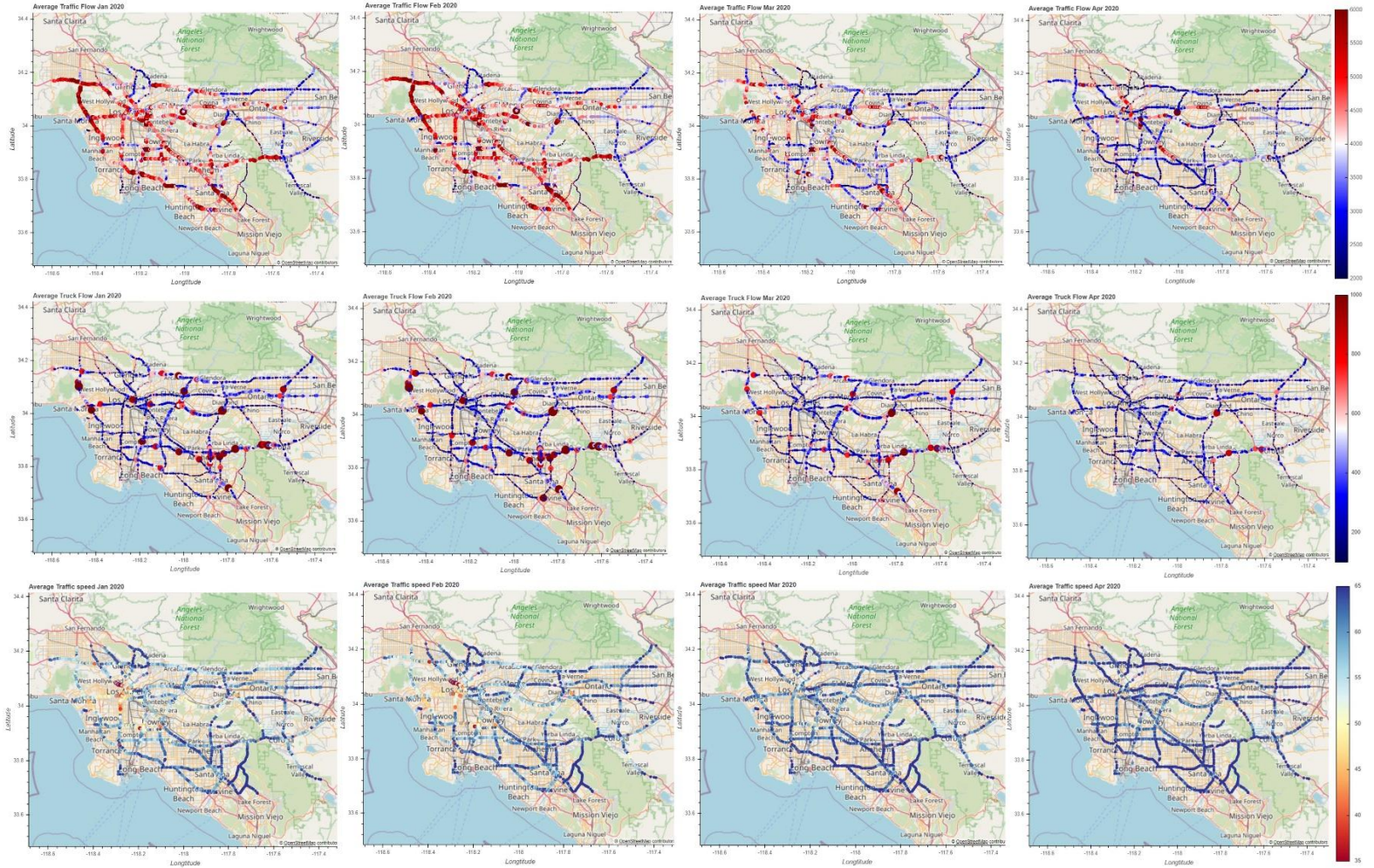

**Fig 10. Spatial pattern for Traffic Flow, Truck Flow and Traffic Speed from Jan to April in 2020, the size of dot represents the value of variables, the bigger the size the larger the value.**

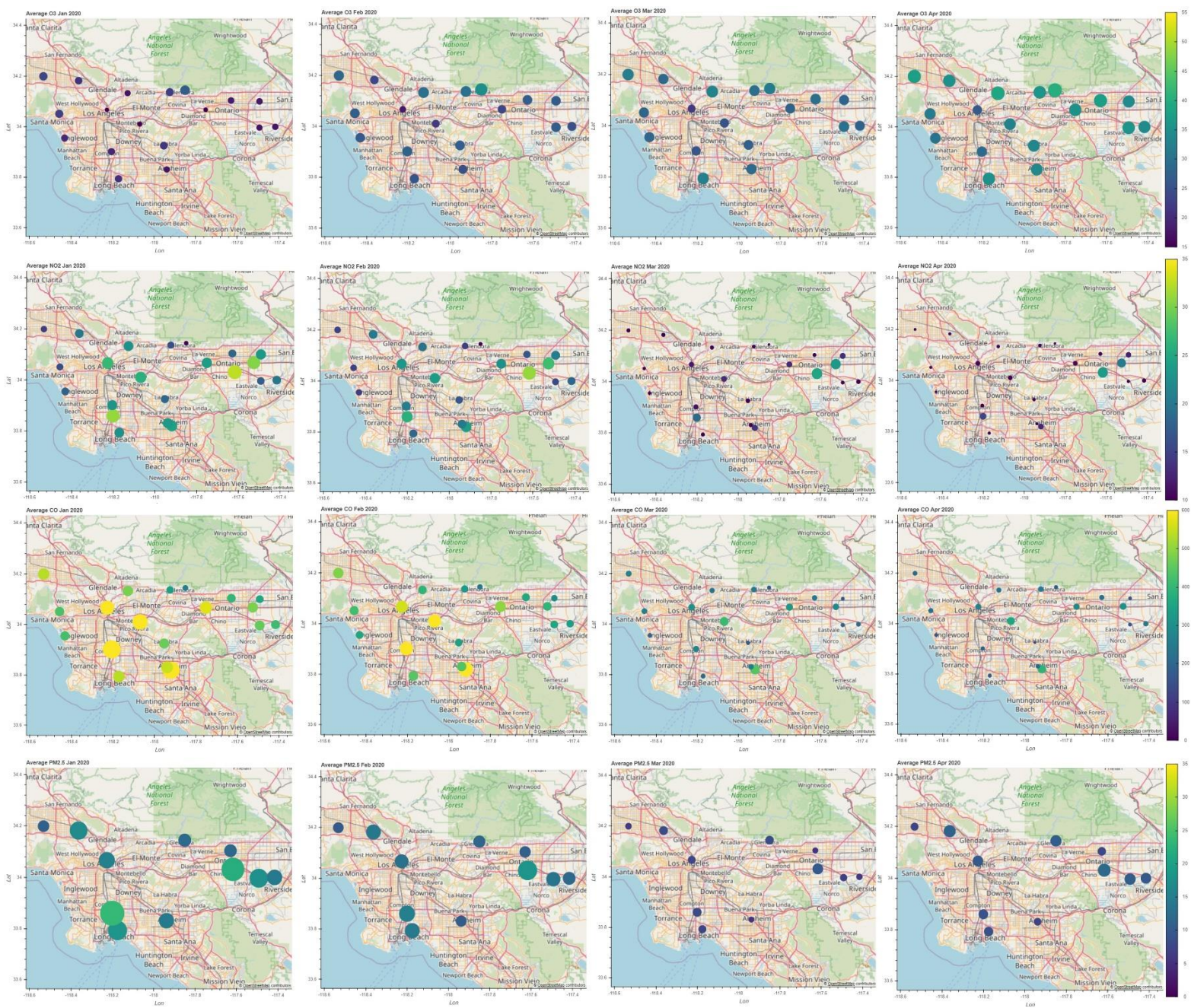

**Fig 11. Spatial pattern for $O_3$, $NO_2$, CO and $PM_{2.5}$ from Jan to April in 2020, the size of dot represents the value of variables, the bigger the size the larger the value.**